\documentclass{aa}
\usepackage{graphicx}

\begin{document}

%   \thesaurus{
%08 (    
%	08.19.4; % {\itshape (Stars:)} supernovae: general 
%        08.03.4; % {\itshape (Stars:)} circumstellar matter
%        02.18.5; % Radiation mechanisms: non-thermal
%        02.18.7; % Radiative transfer
%        08.19.5 SN~1993J % (Stars:) supernovae: individual: 
%   )
%} 
%
   \title{The role of synchrotron self-absorption in the late radio emission
	  from SN~1993J}

%   \subtitle{Late radio emission from SN 1993J}

   \author{M.A.\ P\'erez-Torres 
          \inst{1}
          \fnmsep\thanks{\emph{Present address:} Istituto di Radioastronomia,
                Via P. Gobetti 101, 40129 Bologna, Italy}
          \and A.\ Alberdi\inst{2}
          \and J.M.\ Marcaide\inst{1}
   }

   \offprints{M.A.\ P\'erez-Torres, \email{torres@ira.bo.cnr.it}}

   \institute{Departamento de Astronom\'{\i}a y 
           Astrof\'{\i}sica, Universidad de Valencia, 
           E-46100 Burjassot, Valencia, Spain\\
           email: torres@ira.bo.cnr.it, J.M.Marcaide@uv.es
         \and
	   Instituto de Astrof\'{\i}sica de Andaluc\'{\i}a, CSIC, Apdo.
	   Correos 3004, E-18080 Granada, Spain\\
           email: alberdi@laeff.esa.es
   }

\date{Received: 14 November 2000 / Accepted 23 May 2001}

%
%_________________________________________________________________________	

\abstract{
The standard model for radio supernovae considers 
that the observed synchrotron radio emission 
arises from the high-energy shell
that results from the strong interaction between the expanding 
supernova ejecta and the circumstellar medium.
This emission is considered to be only partially
absorbed by ionized thermal electrons in the circumstellar wind 
of the progenitor star. 
Based on a study of the radio light curves of the type II 
supernova SN1993J, we present evidence of synchrotron 
self-absorption. 
Our modeling of the radio light curves requires 
a large initial magnetic field, of about 30 Gauss, 
and the existence of an
(initially) highly-relativistic population of electrons. 
We also show that while at early epochs the dominant absorption mechanism is 
external absorption by thermal electrons, at late epochs 
and long wavelengths the dominant 
absorption mechanism is synchrotron self-absorption. 
Consequently, the spectral turnover takes place at much shorter 
wavelengths than expected in the standard model, 
and at long wavelengths ($\ge 90$cm at current epochs) the 
flux predictions depart substantially from those 
of the standard model.
    \keywords{radiation mechanisms: non-thermal
                radiative transfer -- 
                supernovae: general --
                supernovae: individual (SN~1993J)
    }   
}

\maketitle

\titlerunning{Synchrotron self-absorption in SN~1993J}
\authorrunning{P\'erez-Torres et al.}

\section{Introduction}
\label{sec:intro}

 Radio emission from supernovae has been previously interpreted in terms
of an optically-thin shell, whose emission is partially suppressed
by external free-free absorption (e.g., Chevalier~\cite{che82a}; 
Weiler et al. \cite{wei89}, \cite{wei90}).
In this model -- known as the standard interaction model 
(Chevalier \cite{che82a}),
hereafter SIM -- a strong interaction between 
the expanding supernova ejecta  
($\rho_{\rm ej} \propto r^{-n}$) and the
circumstellar medium ($\rho_{\rm cs} \propto r^{-s}$)
is expected.
In the SIM, this interaction causes the formation of a self-similarly
expanding ($R_{\rm sh} \propto t^{m}; m = (n-3)/(n-s)$)
shell-like structure (Chevalier \cite{che82b}, Nadyozhin \cite{nad85}), 
from which the observed synchrotron radio emission arises.
In addition, it is assumed (Chevalier \cite{che82a}) that
(i) both magnetic energy density, 
$\epsilon_{\rm B} \propto B^2$, 
and relativistic energy density, 
$\epsilon_{\rm rel} \propto N_0\, E^{2-p}$, 
evolve with time as the post-shock thermal energy density
($\propto \rho_{\rm sh} v_{\rm sh}^2$);
(ii) the synchrotron emission is optically thin;
and (iii) the external absorbing medium has
a power-law dependence $s=2$.

The discovery of SN~1993J (Garc\'{\i}a \cite{gar93}) on March 28, 1993 
in the galaxy M81, 
brought an unprecedented opportunity to test the SIM for supernovae.
The early discovery of hydrogen lines in its spectra 
(Fillipenko \& Matheson \cite{fil93}) classified this supernova as Type II. 
Marcaide and coauthors showed the radio emission 
to arise from a shell-like structure 
(Marcaide et al. \cite{mar95a}) that expands 
self-similarly (Marcaide et al. \cite{mar95b}), 
thus giving strong support to the SIM. 
Marcaide et al. (\cite{mar97}) also reported on the deceleration of the 
expansion ($m=0.86 \pm 0.02$), and concluded that there 
is strong evidence for the circumstellar medium density profile to be
of the form $\rho_{\rm cs} \propto r^{-s}$,  
with $s=1.66$, in good agreement with previous modeling of 
%%both 
radio observations 
($s=1.5$; Van Dyk et al. \cite{van94}, Lundqvist \cite{lun94})  
%%radio and X-ray observations, which require $1.5\, \la s \la\, 1.7$
and with the modeling of both radio and X-ray observations, 
which require $1.5\, \la s \la\, 1.7$
(Fransson et al.~\cite{flc96}; hereafter FLC96).
Attempts to fit the radio light curves 
of SN~1993J using the SIM have been unsuccessful so far,
and have led several authors 
(Van Dyk et al. \cite{van94}, Lundqvist \cite{lun94},
FLC96, Marcaide et al. \cite{mar97}) 
to consider alternative power-law profiles 
for the circumstellar medium around SN~1993J, 
or to include synchrotron self-absorption 
as an additional absorbing mechanism (Fransson \&
Bj\"ornsson \cite{fb98}, hereafter FB98).

\section{Synchrotron radio emission from SN~1993J}
\label{sec:section2}

In an attempt to gain insight into the physical conditions
of SN~1993J, and also motivated by our 
VLBI monitoring of this unique radio supernova,
we have written a numerical code (P\'erez-Torres \cite{per99}) 
that computes the synchrotron radio emission from an expanding 
Type II supernova. 
Although our code will be fully described elsewhere, 
we briefly outline it here.
We solve the synchrotron radiative transfer 
equation throughout the shell (including synchrotron emission 
and synchrotron self-absorption), and take into account both 
synchrotron and expansion losses. 
We assume a (modified) SIM scenario, 
in which the supernova synchrotron radio emission 
comes from the interaction shell that forms shortly after shock breakout. 
(The SIM scenario is modified in the sense that we do not assume the 
synchrotron emission to be optically thin.)  
We assume the emission to be due to the existence of relativistic electrons, 
$N(E,r)$, orbiting in a random magnetic field, $B(r)$. 
We assume an initially injected relativistic electron distribution 
with a power-law energy dependence, $N(E,r) = N_0(r) \, E^{-p}$, 
and also assume a continuous reacceleration of those electrons.

%\subsection{Modeling the radio light curves of SN~1993J}

In modeling the radio emission from SN~1993J,
we take into account all available observational data:
(i) the distance to SN1993J,  $D =$ 3.6 Mpc (Freedman et al. \cite{fre94});
(ii) the supernova explosion date, $t_{\rm exp}$ = March 28.0, 1993  
(Nomoto et al. \cite{nom93}, Podsiadlowski et al. \cite{pod93}, 
Bartunov et al. \cite{bar94});
(iii) the supernova deceleration parameter, $m = 0.86$  
(Marcaide et al. \cite{mar97});
(iv) the width of the radio emitting shell, $\Delta R$ = 0.3\, $R_{\rm shock}$  
(Marcaide et al. \cite{mar95b, mar97}); 
(v) the index of the circumstellar density material, $s = 1.66$ 
(Marcaide et al. \cite{mar97}).
Finally, we assume (vi) the same 
external electron temperature profile as in FB98:
$
T_e (r) = max \left[ T_{15} \left( 10^{15}\, {\rm cm} / r \right) , 
            2 \times 10^5\, {\rm K} \right] 
$,
where $T_{15}$ is the electron temperature in the circumstellar
medium at $r = 10^{15}$ cm, which in our model is equal to 
$1.7 \times 10^6$ K.
The thermal electron density of the circumstellar medium (assumed
to be fully ionized with abundances $X = 0.73, Y = 0.23$), 
is given by 
$
%n_{cs} = 6.34 \times 10^8 \,  M_W \,
n_{cs} = 1.27 \times 10^8 \,  M_W \,
         \left( r / 6.35 \times 10^{14}\, {\rm cm} \right)^{-s} 
	 \, {\rm cm^{-3}}
$,
where we have introduced the wind parameter
$ M_W \equiv 
  \left( \dot{M} / 10^{-5} \, M_\odot\, {\rm yr^{-1}}  \right) \,
  \left( v_w / 10 \, {\rm km\,s^{-1}}  \right)^{-1}  \, {\rm g\,cm^{-1}} 
$, 
and  $\dot{M}$ and $v_w$ are the mass-loss rate and velocity, respectively,
of the presupernova wind.
As in the SIM, we assume that both $B$ and $N_0$ evolve with time 
as the post-shock thermal energy
density, $\rho_{\rm sh} v_{\rm sh}^2$. 
With $m=0.86$ and $s=1.66$ from our VLBI observations, 
we then obtain the following evolution laws for magnetic
field and relativistic particles, respectively:
$ B  \propto  r^{1 - s/2 - 1/m} = r^{-0.99}, 
  N_0 \propto  r^{2 \left( 1 - s/2 - 1/m \right) } = r^{-1.99} 
$, that is, $B \propto r^{-1}$, and $N_0 \propto r^{-2}$.
 These scalings correspond to those in model 4 of 
 Chevalier (\cite{che96}), which give the slowest evolution of 
 $B$ and $N_0$, and therefore of the radio emission from SN1993J.
 The other three models consider a dependence of $B \propto t^{-1}$,
or $N_0 \propto t^{-2}$ or both.

\begin{figure}
        \resizebox{\hsize}{!}{\includegraphics{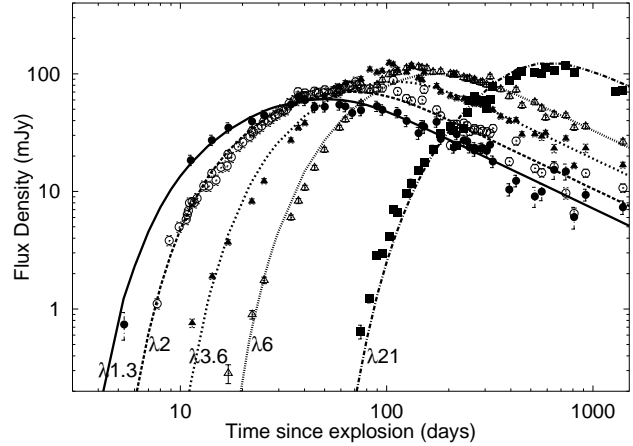}}
        \caption[./h2783f1.eps]{
		Radio light curves for SN~1993J at
 		$\lambda$1.3 cm (filled circles),
 		$\lambda$2 cm (open circles),
 		$\lambda$3.6 cm (filled triangles),
	        $\lambda$6 cm (open triangles), and
 		$\lambda$21 cm (filled squares),
		from measurements made using the VLA 
                (Van Dyk et al.~\cite{van94}, and Van Dyk, private communication) 
                and, at 2 cm,
		also the Ryle Telescope (Pooley \& Green~\cite{poo93}).
 		The lines represent our best fit model as described in the
 		text. 
              }
         \label{fig:fig1}
   \end{figure}

We model the available radio data using the following parameters: 
(1) the spectral index of the injected electron distribution, $p$;
(2) the initial magnetic field, $B (r_0)$;
(3) the initial value of the injected function of relativistic electrons, 
$N_0(r_0)$;
and (4) a low-energy cut-off for the relativistic electrons, 
$\gamma_{\rm min}(r_0) \equiv E_{\rm min}(r_0)/m_e\,c^2$. 
We fit the above parameters at the reference radius $r_0 = 6.35 \times 
10^{14}$\,cm, corresponding to the outer shell radius at our first epoch 
($t_0 = 2.34 $ days).
We find that the multi-wavelength radio data for SN~1993J 
(Figure~\ref{fig:fig1}) are best reproduced by a model with 
$p = 3$, 
$B (r_0) = 27$ G, 
$N_0(r_0) = 6.7\times 10^{-7} {\rm erg^{p - 1}\, cm^{-3}}$,
and $\gamma_{\rm min}(r_0) \approx 90$.

%                                                One column figure
%-------------------------------------------------------------------------
%

%
%-------------------------------------------------------------------------
%

We estimate the uncertainties in the parameters of our model by modifying 
each of the best-fit parameters (one at a time), 
and comparing via a least-squares fit the 
resultant emission with that predicted by our best-fit model.  
These calculations show the extreme sensitivity of 
our results to $p$, which needs to be kept very close to 3 
(within a small few per cent) to obtain any meaningful fit to the 
radio light curves.   
The other three parameters are less well constrained, and a 20\,\%  variation
of each of their best-fit values result in reasonable fits.  
In addition, we have considered the existence of 
couplings between some of the parameters, especially  $N_0$ and $B$. 
Further testing shows that factor of two variations 
in each of the three poorly-constrained parameters 
give very poor fits to the observations. 
Therefore, we adopt factor of two modeling errors
for each of those three  parameters. 

\subsection{Spectral evolution of SN1993J: evidence for synchrotron 
self-absorption}

\begin{figure}
      \resizebox{\hsize}{!}{\includegraphics{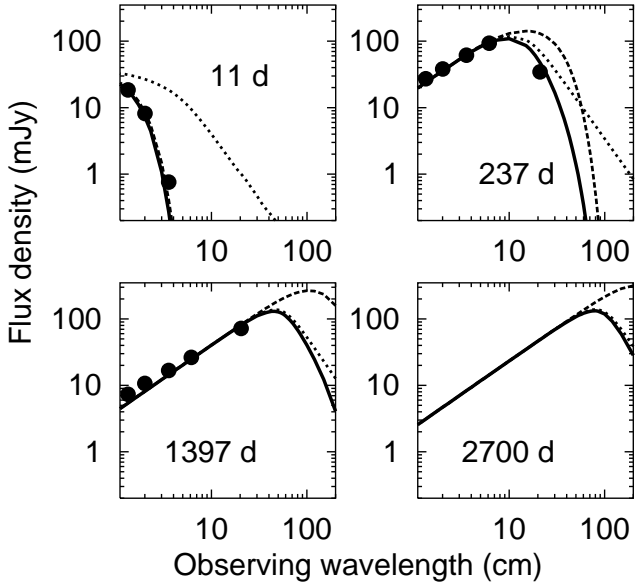}}
      \vspace{45pt}
      \caption[h2783f2.eps]{
	Fits to spectra at 11, 237, and 1397 days after explosion,
	and predicted spectrum at 2700 days,
	considering external free-free absorption (dashed lines),
	synchrotron self-absorption (dotted lines),
	and both external free-free absorption
	and synchrotron self-absorption (solid lines).
	The  filled circles represent  the
	available observational data for each epoch.
	At early epochs, the radio spectrum of SN~1993J (upper left plot)
	is well explained by the solely effect of free-free absorption from the
	external medium.
	At late epochs (lower left plot), however,
	both free-free absorption and synchrotron
	self-absorption predict a similar spectral behavior for the observed
	wavelength range, but a different one at wavelengths around and above
	$\lambda 40$ cm, at which observations are not available.
        The lower right-hand plot shows the predicted spectrum at current 
        epochs.
	This prediction can be tested to discriminate
	between the (external) free-free and (internal) 
	self-absorption mechanisms described in the text.
        }
      \label{fig:fig2}
   \end{figure}

Our simulated spectra for SN~1993J agree well with the observed 
spectra (Fig.~\ref{fig:fig2}).   
The supernova spectrum changes from a mainly inverted spectrum at early 
epochs to a more typical optically thin 
spectrum at late epochs and for an increasingly wide 
wavelength range. 
As shown in Fig.~\ref{fig:fig2}, external free-free absorption 
is dominant at early times ($t \leq 100$ days), 
but synchrotron self-absorption becomes 
increasingly important with time for the long-wavelength 
radio emission.  
Thus, the standard interaction model is incomplete, at least for SN~1993J, 
in that it ignores synchrotron self-absorption, which according to our 
modeling is the most relevant absorption mechanism -- as the supernova ages -- 
for the longer wavelengths.  
At current epochs, our model predicts that the flux at $\lambda 90$cm 
is $\sim$120 mJy for the synchrotron self-absorption dominated scenario, 
and $\sim$200 mJy for the free-free absorption dominated scenario. 
This prediction -- which would represent a direct test  
to discern the relevant absorption mechanism for SN~1993J --
can be checked using $\lambda \sim$90 cm observations of SN~1993J with a
sensitive radio interferometer. 

%                                                
%--------------------------  FIGURE      -------------------------------------
%
%                                                
%-------------------------- END OF FIGURE -------------------------------------
%

\section{Results and Discussion}

\begin{figure}
        \resizebox{\hsize}{!}{\includegraphics{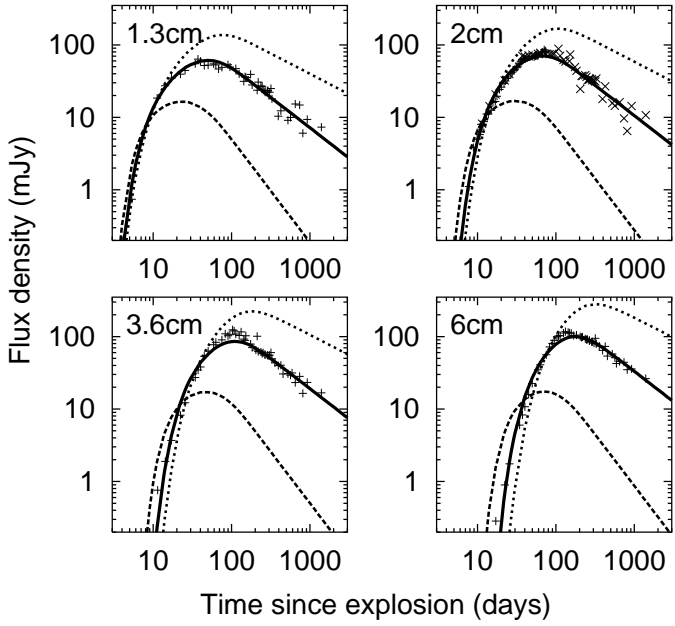}}
        \vspace{70pt}
        \caption[h2783f3.eps]{
 	Observed radio light curves for SN~1993J at 1.3, 2, 3.6, and 6cm, 
        and model predictions for three different
	indices of the presupernova circumstellar wind: 
        $s=1.5$ (dotted lines), $s=1.66$ (solid lines), 
        $s=2$ (dashed lines).  See text for details.
        }
         \label{fig:fig3}
   \end{figure}

Our modeling results show that a combination of synchrotron self-absorption
and external free-free absorption can reproduce well the 
evolution of the radio light curves of SN~1993J.
Therefore, the {\sl ad hoc} external inhomogeneous medium 
(``clumpy external absorption'') previously invoked by 
Van Dyk et al. (\cite{van94}) to model the radio light curves 
seems to be unnecessary.
We have also considered the Razin-Tsytovich effect 
(the suppression of the emission below a certain wavelength due to the 
existence of a plasma) as an alternative mechanism to 
free-free absorption.
However, the densities of thermal electrons required for the 
Razin-Tsytovich mechanism to be effective are unrealistically large. 
For example, at 11.5 days, we find that 
the number density of thermal electrons should be larger than  
$\sim 1.3 \times 10^9 {\rm cm^{-3}}$ for the Razin-Tsytovich 
effect to be important.  This value is about an order of magnitude
larger than needed to correctly model the X-ray
luminosity at those epochs (FLC96). 
The same holds at late epochs, and we therefore rule out the
Razin-Tsytovich effect as a plausible alternative to 
external free-free absorption.
We point out here that the index of the circumstellar 
medium density profile,  
$s=1.66$, was obtained by Marcaide et al.~(\cite{mar97})
under the assumption that external free-free absorption 
was the only absorption mechanism  acting
in SN1993J. Although there is increasing evidence that
synchrotron self-absorption is also relevant for
SN~1993J (Chevalier~\cite{che98}, FB98, this work), we
note that at the early epochs used by Marcaide et al. (\cite{mar97})
free-free absorption is the dominant mechanism, while 
synchrotron self-absorption can be neglected (see Figure
2, epoch 11 days). We have tried to fit
the light curves using either an standard
index for the circumstellar wind ($s=2$), or a much flatter one
($s=1.5$) which would still satisfy the range of $1.5\, \la s \la\, 1.7$
inferred from X-ray observations (FLC96). 
As Fig.~\ref{fig:fig3} shows,  an $s=2$ wind causes a fall-off of 
the observed emission which is far too rapid.  
On the contrary, a wind with $s=1.5$ makes the 
radioemission to fall-off too slowly. 
For the case $s=2$, we also investigated different
scalings for $B$ and $N_0$; 
namely, we used the other three scalings mentioned by 
Chevalier(\cite{che96}). 
However, none of those scalings gives good fits because
all of them make both $B$ and $N_0$ --and therefore the radio
light curves-- to decrease more rapidly than $r^{-1}$ and $r^{-2}$, respectively.
This suggests that the index $s$ should be close 
to our nominal value of $s=1.66$. 
Since the decline rate of the radio light curves, i.e., their
late time emission, is also affected by the parameter $p$, we 
have investigated models in which both $s$ and $p$ are allowed to vary.
This search resulted in much poorer fits than our best one. 
We acknowledge, though, that the assumption of {\em ad hoc} scalings for 
both $B$ and $N_0$ could alleviate these poor fits, at the expense of more complicated
models. We defer such detailed study to a further publication. 
%where all the details of the code will be explained.}

We investigated as well the effects of using different 
indices for the expansion ($m$), but 
do not find a reasonable fit to the supernova emission
using either an undecelerated expansion of the supernova 
($m = 1$), or a heavily decelerated expansion ($m = 0.75$). 

%                                                One column figure
%-------------------------------------------------------------------------
%

%
%-------------------------------------------------------------------------
%

The unexpected large (initial) magnetic 
field of about $30$G implied by our modeling 
agrees with the large magnetic field obtained by FB98, 
and seems to rule out any hypothesis based on a magnetic field 
due to a compression of the circumstellar wind field.  
Indeed, since it is very unlikely that the magnetic field energy density in the 
wind is larger than its kinetic energy density, i.e., 
$B^2/8\pi \leq \rho v^2_{\rm w}/2$, we should expect a magnetic field not 
in excess of $B \leq$ 3.4 mG at $r \sim 3 \times 10^{16}$cm (if
it arises from a compression of the magnetic field in the wind). 
This value of the magnetic field is in agreement with the values obtained 
from polarization observations of OH masers in supergiants 
(e.g., Cohen et al.~\cite{cohen87}, Nedoluha \& Bowers~\cite{nedo92}), 
which give magnetic field estimates at distances of $\sim 3 \times 10^{16}$cm of
$\sim 1-10$ mG.  
However, we find that the magnetic field is actually about 
0.58 G, i.e., 170 times larger.
From a strong supernova shock (factor of 4 compression), 
one would expect an increase of the field value of $\sim 4^{1/s}$ 
for a magnetic field that scales as $B \propto r^{-1} \propto n_{\rm sh}^{1/s}$,
clearly insufficient by almost two orders of magnitude.
Therefore, the large magnetic field   
argues strongly in favor of turbulent amplification in the 
shell (e.g., Gull \cite{gul73}, Chevalier \cite{che82a},
Chevalier \& Blondin \cite{che95}). 
%Understanding in detail the origin of these strong magnetic fields
%will surely be a challenging new avenue of theoretical work, 
%but is beyond the scope of this paper.
Furthermore, the assumption that the magnetic energy density evolves 
as the post-shock thermal energy results in a dependence 
of the magnetic field with radius of $r^{-1}$. 
A component of the field behaving like $B \propto r^{-2}$, 
as suggested from magnetic flux conservation, does not give a 
reasonably good fit to the observations. 

We find that the magnetic energy density in the shell is just
a (small) fraction of the post-shock thermal energy density,
$\epsilon_B / \epsilon_{\rm sh} \approx 0.024$, but nonetheless
much larger than the relativistic electron energy density, 
$\epsilon_{\rm rel} / \epsilon_{\rm sh} \approx 2 \times 10^{-4} - 10^{-3}$. 
This points to a strong deviation from equipartition. 
We note, however, that the ions (whose emission we neglect as it
is a factor $m_p / m_e \approx 2000$ less than that due 
to the electrons) might have a large contribution to the total
energy density.
We also find that the relativistic particle number density is a 
very small fraction of the particle thermal density. 
In fact, the ratio 
$ n_{\rm rel}/n_{\rm sh} = 
     (N_0\, E_{\rm min}^{1-p}/2\,n_{\rm sh}) \approx (1 - 8) \times 10^{-7}$
at any given epoch.

\begin{figure}
        \resizebox{\hsize}{!}{\includegraphics{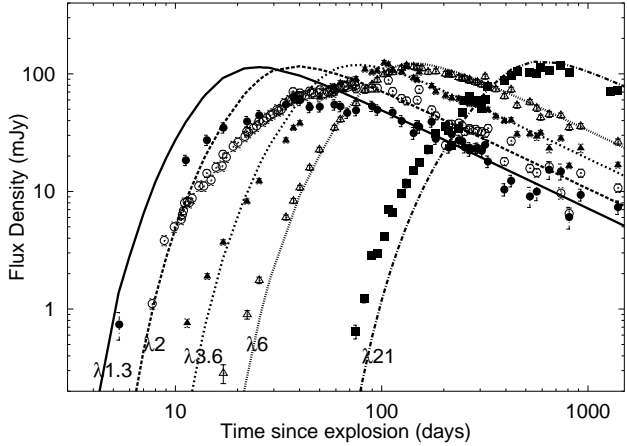}}
        \caption[./h2783f4.eps]{
	Radio light curves for SN~1993J. 
	The data is as in Fig.~\ref{fig:fig1}. 
        The lines correspond to our best-fit model as described in  the
	text, except that we did not assume a low-energy cut-off. 
        Such model fails to reproduce the flattening of the short-wavelength
	emission seen at early epochs. See text for details.
              }
         \label{fig:fig4}
   \end{figure}

 Special attention deserves the low-energy cutoff at early epochs
that we consider to account for the flattening of the radio 
light curves at early epochs. 
We recall that our model assumes an initial injection
of (highly) relativistic particles $N(E,r) = N_0(r) \, E^{-p}$,
which applies from $E_{min}$ up to $E_{max}$. 
The initial supernova explosion is the likely seed for 
such a population of relativistic electrons, 
which must be continuously reaccelerated.
We show in Fig.~\ref{fig:fig4} the light curves
that we obtain when we do not consider a low-energy
cutoff in the relativistic particle distribution, 
i.e., $\gamma_{\rm min} = 1$.  
(The other parameters are fixed to their nominal values
$p = 3$, $B (r_0) = 27$ G, and
$N_0(r_0) = 6.7\times 10^{-7} {\rm erg^{p - 1}\, cm^{-3}}$.)
We note that although the optically thin part of the radio 
light curves remains unaffected, the optically thick part 
is heavily modified by the lack of a low-energy
cutoff. In particular, the radio emission rises more rapidly. 
More importantly, the observed flattening of the light curves
at the shortest wavelengths is not reproduced by such a model.
Notice also that at long wavelengths 
(see the light curve at $\lambda21$cm in Fig.~\ref{fig:fig4}) 
the radio emission is strongly absorbed at early times.  
This is the result of having decreased the minimum energy
of the electrons; 
since the number of relativistic particles is proportional
to $E_{\rm min}^{1-p}$, we effectively increased the column density
of the relativistic particles by a large factor. 
The effect is most easily seen at long wavelengths due to the
dependence of the opacity with wavelength  
($\kappa_{\nu} \propto \lambda^{\rm (p+4)/2}$).
At this point, we also note 
that our model differs here from that presented
by FB98, in that they assume a continuous injection of particles
with $\gamma \geq 1$.  
Bell~(\cite{bell78a}) showed that particles that 
are {\em already} relativistic can be accelerated in the presence of a shock.
Since it is usually assumed that the 
acceleration of particles in SNe is due to the Fermi process (e.g.
Ball \& Kirk~\cite{bal95} for SN1986J), we show here that our finding
of a low-energy cutoff is consistent with such a process acting in SN1993J. 
The minimum energy of electrons that can be efficiently accelerated is 
$\gamma_{\rm min, e} = (m_p/m_e)(v_{\rm A}/c\mu)$ 
(Eilek \& Hughes~\cite{eilek91}; 
but see also Ellison et al.~\cite{ellison2000},
who argue that nonlinear effects in shocks can lead to acceleration of  
particles from a thermal pool), 
where $m_p$ and $m_p$ are the protron and electron mass,
respectively, $v_{\rm A}$ is the Alfv\'en velocity, 
$c$ is the speed of light, 
and $\mu$ is the cosine of the pitch-angle. 
With the physical conditions deduced for SN 1993J,
$v_{\rm A}$ varies from $\approx 3200$ km s$^{-1}$ at t=2.3 days
down to about 1200 km s$^{-1}$ at t=3000 days. This gives
$\gamma_{\rm min} > 20$ and  $\gamma_{\rm min} > 7$ at 2.3 
and 3000 days, respectively, which is consistent with our results.

%                                                One column figure
%-------------------------------------------------------------------------
%

%
%-------------------------------------------------------------------------
%

It is plausible that Coulomb losses also play a role in 
the shape of the SN~1993J light curves at  high frequencies
and early epochs (see FB98, which use power-laws 
for the circumstellar wind density and the deceleration of the 
expansion of SN~1993J that are different from the experimental values 
obtained by Marcaide et al.~\cite{mar97}). 
However, the neglect of Coulomb losses in our modeling is self-consistent,
and does not affect the behavior of the radio light curves 
either at early or at late epochs. 
Indeed, the Lorentz factor, $\gamma$, of the electrons responsible for
the emission at a given wavelength is 
$\gamma \approx 
       84\, \left( \lambda\, B\, \sin \theta \right)^{-1/2}$ 
(cgs units), 
while the Lorentz factor below which Coulomb losses dominate
over expansion losses is approximately 
$\gamma_{\rm Coul} \approx 
       180 \, \left( n_{\rm sh}/ 6.9 \times 10^7 {\rm cm^{-3}} \right)\,
             \left( t / 10 {\rm days} \right)^{1 - m \, s}$, 
where we have neglected for simplicity the slowly-dependent term on
the electron energy.
From our modeling at $t = 10$ days, we get
$B \approx 7.7$G and $\gamma_{\rm min} \approx 61$, so
the observed emission at $\lambda$1.3 cm comes mainly from electrons
with $\gamma \approx 200$, while Coulomb losses are important
for $\gamma \la \gamma_{\rm Coul} \approx 180$.  
Analogously, our modeling gives a 
magnetic field of $\sim$ 65 mG at current epochs ($\sim$2700 days), 
and $\gamma_{\rm min} \approx 12$.  
We then obtain that the observed $\lambda$90 cm emission
comes mainly from electrons with $\gamma \approx$120, 
while Coulomb losses are relevant only for electrons with 
$\gamma \la \gamma_{Coul} \approx 17$, 
and thus do not affect the long wavelength radio emission from
SN~1993J at late epochs.
Therefore, we stress that our main result,  namely the dominance
of synchrotron self-absorption over external free-free absorption
at late epochs is a robust result, and is not affected by the 
physical details that cause the flattening of the spectra at early epochs.

In summary, we have presented evidence of synchrotron
self-absorption for the type II SN~1993J, based 
on a modeling of its radio light curves. 
We show that the radio light curves 
of SN~1993J are better characterized by synchrotron emission
that is partially absorbed by a combination of synchrotron
self-absorption (inside the shell) and external free-free absorption, 
rather than only by external free-free absorption, 
thus confirming previous modeling of the SN~1993J radio light
curves (FB98).
This result suggests that the ``clumpy external absorption'' 
previously invoked to model the radio light curves of SN~1993J
might not be needed.
We also find that the Razin-Tsytovich effect can be
ruled out as well, as an alternative mechanism to external free-free absorption.
We predict that synchrotron self-absorption is the 
main absorbing mechanism for the long-wavelength radio emission 
as SN~1993J ages. 
This prediction can be tested unambiguously 
using current technology and observations of SN~1993J at $\lambda 90$ cm. 
Moreover, follow-up observations of SN~1993J at $\lambda \geq 90$ cm 
are also needed to further constrain the physical parameters
of this unique supernova.  
As pointed out earlier (Chevalier~\cite{che98}),
if SN~1993J is not a bizarre supernova 
then synchrotron self-absorption is likely to play an important
role in most, if not all, radio supernovae, especially as they age.

\begin{acknowledgements}
We thank an anonymous referee for useful comments on the manuscript.
This work has been partially supported by the
 Spanish DGICYT Grants No. PB96-0782 and PB97-1164.
\end{acknowledgements}

\end{document}